\documentstyle[twoside,epsf]{article}

\catcode`\@=11
\long\def\@makefntext#1{
\protect\noindent \hbox to 3.2pt {\hskip-.9pt
$^{{\eightrm\@thefnmark}}$\hfil}#1\hfill}		%CAN BE USED

\def\thefootnote{\fnsymbol{footnote}}
\def\@makefnmark{\hbox to 0pt{$^{\@thefnmark}$\hss}}	%ORIGINAL

\def\ps@myheadings{\let\@mkboth\@gobbletwo
\def\@oddhead{\hbox{}
\rightmark\hfil\eightrm\thepage}
\def\@oddfoot{}\def\@evenhead{\eightrm\thepage\hfil
\leftmark\hbox{}}\def\@evenfoot{}
\def\sectionmark##1{}\def\subsectionmark##1{}}

\oddsidemargin=\evensidemargin
\renewcommand{\thefootnote}{\fnsymbol{footnote}}

\newcounter{sectionc}\newcounter{subsectionc}\newcounter{subsubsectionc}
\renewcommand{\section}[1] {\vspace{12pt}\addtocounter{sectionc}{1}
\setcounter{subsectionc}{0}\setcounter{subsubsectionc}{0}\noindent
	{\tenbf\thesectionc. #1}\par\vspace{5pt}}
\renewcommand{\subsection}[1] {\vspace{12pt}\addtocounter{subsectionc}{1}
	\setcounter{subsubsectionc}{0}\noindent
	{\bf\thesectionc.\thesubsectionc. {\kern1pt \bfit #1}}\par\vspace{5pt}}
\renewcommand{\subsubsection}[1] {\vspace{12pt}\addtocounter{subsubsectionc}{1}
	\noindent{\tenrm\thesectionc.\thesubsectionc.\thesubsubsectionc.
	{\kern1pt \tenit #1}}\par\vspace{5pt}}
\newcommand{\nonumsection}[1] {\vspace{12pt}\noindent{\tenbf #1}
	\par\vspace{5pt}}

\newcounter{appendixc}
\newcounter{subappendixc}[appendixc]
\newcounter{subsubappendixc}[subappendixc]
\renewcommand{\thesubappendixc}{\Alph{appendixc}.\arabic{subappendixc}}
\renewcommand{\thesubsubappendixc}
	{\Alph{appendixc}.\arabic{subappendixc}.\arabic{subsubappendixc}}

\renewcommand{\appendix}[1] {\vspace{12pt}
        \refstepcounter{appendixc}
        \setcounter{figure}{0}
        \setcounter{table}{0}
        \setcounter{lemma}{0}
        \setcounter{theorem}{0}
        \setcounter{corollary}{0}
        \setcounter{definition}{0}
        \setcounter{equation}{0}
        \renewcommand{\thefigure}{\Alph{appendixc}.\arabic{figure}}
        \renewcommand{\thetable}{\Alph{appendixc}.\arabic{table}}
        \renewcommand{\theappendixc}{\Alph{appendixc}}
        \renewcommand{\thelemma}{\Alph{appendixc}.\arabic{lemma}}
        \renewcommand{\thetheorem}{\Alph{appendixc}.\arabic{theorem}}
        \renewcommand{\thedefinition}{\Alph{appendixc}.\arabic{definition}}
        \renewcommand{\thecorollary}{\Alph{appendixc}.\arabic{corollary}}
        \renewcommand{\theequation}{\Alph{appendixc}.\arabic{equation}}
        \noindent{\tenbf Appendix \theappendixc #1}\par\vspace{5pt}}
\newcommand{\subappendix}[1] {\vspace{12pt}
        \refstepcounter{subappendixc}
        \noindent{\bf Appendix \thesubappendixc. {\kern1pt \bfit #1}}
	\par\vspace{5pt}}
\newcommand{\subsubappendix}[1] {\vspace{12pt}
        \refstepcounter{subsubappendixc}
        \noindent{\rm Appendix \thesubsubappendixc. {\kern1pt \tenit #1}}
	\par\vspace{5pt}}

\newcommand{\textlineskip}{\baselineskip=13pt}
\newcommand{\smalllineskip}{\baselineskip=10pt}

\def\eightcirc{
\begin{picture}(0,0)
\put(4.4,1.8){\circle{6.5}}
\end{picture}}
\def\eightcopyright{\eightcirc\kern2.7pt\hbox{\eightrm c}}

\def\abstracts#1#2#3{{
	\centering{\begin{minipage}{4.5in}\baselineskip=10pt\footnotesize
	\parindent=0pt #1\par
	\parindent=15pt #2\par
	\parindent=15pt #3
	\end{minipage}}\par}}

\renewenvironment{thebibliography}[1]
	{\frenchspacing
	 \ninerm\baselineskip=11pt
	 \begin{list}{\arabic{enumi}.}
	{\usecounter{enumi}\setlength{\parsep}{0pt}
	 \setlength{\leftmargin 12.7pt}{\rightmargin 0pt} %FOR 1--9 ITEMS
	 \setlength{\itemsep}{0pt} \settowidth
	{\labelwidth}{#1.}\sloppy}}{\end{list}}

\newcounter{itemlistc}
\newcounter{romanlistc}
\newcounter{alphlistc}
\newcounter{arabiclistc}

\newcommand{\fcaption}[1]{
        \refstepcounter{figure}
        \setbox\@tempboxa = \hbox{\footnotesize Fig.~\thefigure. #1}
        \ifdim \wd\@tempboxa > 5in
           {\begin{center}
        \parbox{5in}{\footnotesize\smalllineskip Fig.~\thefigure. #1}
            \end{center}}
        \else
             {\begin{center}
             {\footnotesize Fig.~\thefigure. #1}
              \end{center}}
        \fi}

\newcommand{\tcaption}[1]{
        \refstepcounter{table}
        \setbox\@tempboxa = \hbox{\footnotesize Table~\thetable. #1}
        \ifdim \wd\@tempboxa > 5in
           {\begin{center}
        \parbox{5in}{\footnotesize\smalllineskip Table~\thetable. #1}
            \end{center}}
        \else
             {\begin{center}
             {\footnotesize Table~\thetable. #1}
              \end{center}}
        \fi}

\def\@citex[#1]#2{\if@filesw\immediate\write\@auxout
	{\string\citation{#2}}\fi
\def\@citea{}\@cite{\@for\@citeb:=#2\do
	{\@citea\def\@citea{,}\@ifundefined
	{b@\@citeb}{{\bf ?}\@warning
	{Citation `\@citeb' on page \thepage \space undefined}}
	{\csname b@\@citeb\endcsname}}}{#1}}

\newif\if@cghi
\def\cite{\@cghitrue\@ifnextchar [{\@tempswatrue
	\@citex}{\@tempswafalse\@citex[]}}
\def\citelow{\@cghifalse\@ifnextchar [{\@tempswatrue
	\@citex}{\@tempswafalse\@citex[]}}
\def\@cite#1#2{{$\null^{#1}$\if@tempswa\typeout
	{IJCGA warning: optional citation argument
	ignored: `#2'} \fi}}

\def\pmb#1{\setbox0=\hbox{#1}
	\kern-.025em\copy0\kern-\wd0
	\kern.05em\copy0\kern-\wd0
	\kern-.025em\raise.0433em\box0}

\def\fnt#1#2{\footnotetext{\kern-.3em
	{$^{\mbox{\scriptsize #1}}$}{#2}}}

\def\fpage#1{\begingroup
\voffset=.3in
\thispagestyle{empty}\begin{table}[b]\centerline{\footnotesize #1}
	\end{table}\endgroup}

\font\tenrm=cmr10
\font\tenit=cmti10
\font\tenbf=cmbx10
\font\bfit=cmbxti10 at 10pt
\font\ninerm=cmr9

\font\eightrm=cmr8

\newcommand{\beq}{\begin{equation}}
\newcommand{\eeq}{\end{equation}}
\newcommand{\beqa}{\begin{eqnarray}}
\newcommand{\eeqa}{\end{eqnarray}}

\def\qed{\hbox{${\vcenter{\vbox{			%HOLLOW SQUARE
   \hrule height 0.4pt\hbox{\vrule width 0.4pt height 6pt
   \kern5pt\vrule width 0.4pt}\hrule height 0.4pt}}}$}}

\renewcommand{\thefootnote}{\fnsymbol{footnote}}	%USE SYMBOLIC FOOTNOTE
\begin{document}

\def\case#1#2{\textstyle{#1\over#2}}

\fpage{1}

\begin{flushright}
UCSD-96-22\\
September, 1996
\end{flushright}

\vspace{1cm}
\centerline{\bf  HADRO-PRODUCTION OF QUARKONIA}
\vspace*{0.035truein}
\centerline{\bf IN FIXED TARGET EXPERIMENTS \footnote{To appear in the
proceedings of the Quarkonium Physics Workshop, University of Illinois,
Chicago, June 13--15, 1996.}}
\vspace*{0.37truein}
\centerline{\footnotesize IRA Z. ROTHSTEIN}
\vspace*{0.015truein}
\centerline{\footnotesize\it Department of Physics and Astronomy, University of
California, San Diego}
\baselineskip=10pt
\centerline{\footnotesize\it  9500 Gilman Drive, La Jolla, CA 92093, U.S.A}

\vspace*{0.21truein}
\abstracts{In this talk I review the recent progress made in the calculations
of quarkonia production in fixed target experiments. 
NRQCD  organizes the calculations in
a systematic expansion in $\alpha_s$ and $v$, the relative velocity
between the heavy quarks. Within this formalism there are octet
contributions which are not included in the color singlet model. 
These contributions depend upon unknown matrix elements
of local operators which are fit to the data.
Using these fits, there are several predictions which do indeed
improve agreement with the data.
However, the prediction for the
polarization of the produced states as well as the ratio of the
$\chi_1$ to $\chi_2$ cross sections differ substantially from
the data for the case of pion beams. 
Possible large corrections from higher twist effects
are discussed as is the issue of the the proper choice of masses.}{}{}

\textlineskip
\vspace*{12pt}
\textheight=7.8truein
\setcounter{footnote}{0}
\renewcommand{\thefootnote}{\alph{footnote}}
\section{Introduction}

The use of non-relativistic QCD (NRQCD) allows us to calculate both
production and annihilation rates of heavy quark bound states in
 a systematic expansion in  $\alpha_s$ and $v$, the relative 
velocity of the heavy quarks\cite{BOD95}.
Moreover, the inclusion of higher Fock states, which emerge naturally
in the formalism, allows for a consistent factorization of long
and short distant effects, thus validating the
use perturbative QCD to calculate the Wilson coefficients. 
Furthermore, the long distance effects are now
written in terms of well defined operators, which can be calculated on
the lattice \cite{bod}, instead of potential model wave functions. 

In light of this progress, it is interesting to revisit \cite{BRI,GUP96,VAE95} 
the issue of
hadro-production in fixed target experiments. Previous calculations
within the confines of the color singlet model \cite{SCH94} were found to be 
inconsistent with the data\cite{VAE95}. The prediction for the
overall normalization of the cross section  is too small as is the ratio
of the production cross sections for $\chi_1$ and $\chi_2$.
The direct production rate for $J/\psi$ is too small, and the
$J/\psi$  are predicted to be partially transversely polarized, 
which they are not. This disparity between theory and data is, at least for
the first two observables discussed, 
crying out for a new production channel. NRQCD supplies just such a  
channel, namely the state in which the two heavy quarks are in a 
relative octet configuration. This state will have a finite overlap after
soft gluon emission acts as a color sink.
The overlap will be suppressed by powers of $v$ as dictated by the
velocity scaling rules \cite{horn}.

Before proceeding to the results I would like to briefly discuss
the levels of rigor which go into the various approximations 
in the calculations. First, there is no operator product expansion
in these calculations. Thus, the factorization is performed
via a diagrammatic analysis, as is done for Drell-Yan and other
such processes. Factorization in such cases in known to be
violated by higher twist effects which are suppressed by
powers of the large invariant mass scale involved in the process. 
For the case discussed here, this scale  would
be quarkonium mass. In this regard, the proofs of factorization in
the case of
small $p_T$ production is no less rigorous  than at large $p_T$. The
only difference in the two cases is that the higher twist corrections at
large $p_T$ are suppressed by $1/p_T^2$ as opposed to 1/$m_H^2$.
 Thus, we expect larger errors to be incurred at small $p_T$.

 In showing
factorization it is imperative that one sum over all relative
color states of the quarkonium, otherwise, as was discussed in the
case of the decay of P wave states \cite{BBLII}, there will be an infrared
divergent Wilson coefficient which obviously destroys any hope
of calculating in a model independent fashion. Once the higher Fock
states are included, the factorization is restored and the final
result of our calculation depends upon unknown non-perturbative
matrix elements which are enumerated by the velocity scaling rules.
Thus, the production cross section for the reactions
\begin{equation}
\label{proc}
A + B \longrightarrow H + X,
\end{equation}

\noindent can be written as
\begin{equation}
\label{fact}
\sigma_H = \sum_{i,j}\int\limits_0^1 d x_1 d x_2\,
f_{i/A}(x_1) f_{j/B}(x_2)\,\hat{\sigma}(ij\rightarrow H)\,,
\end{equation}
\begin{equation}
\label{factformula}
\hat{\sigma}(ij\rightarrow H) = \sum_n C^{ij}_{\bar{Q} Q[n]} 
\langle {\cal O}^H_n\rangle\,.
\end{equation}

\noindent Here the first sum extends over all partons in the colliding 
hadrons and $f_{i/A}$ etc. denote the corresponding distribution 
functions. The short-distance ($x\sim 1/m_Q \gg 1/(m_Q v)$) 
coefficients $C^{ij}_{\bar{Q} Q[n]}$ 
describe the production of a quark-antiquark pair 
in a state $n$ and have expansions in $\alpha_s(2 m_Q)$. The 
parameters\footnote{Their precise definition is given in 
Sect.~VI of \cite{BOD95}.} 
$\langle {\cal O}^H_n\rangle$ describe the subsequent 
hadronization of the $Q\bar{Q}$ pair into a jet containing 
the quarkonium $H$ and light hadrons.

The velocity scaling 
rules are derived via the multipole expansion, which tells
us that soft gluon couplings to a heavy quark bound state are
suppressed by the ratio of the size of the bound state to
the wavelength of the gluon. This expansion was first used within
the confines of the strong interaction by Gottfried \cite{got}
a while back, and it seems to work quite well. Furthermore, present
extractions of matrix elements seem to agree with  their
predicted scalings with $v$. As such, we will continue under the
assumption that these scaling rules are valid.
Furthermore, we will ignore effects due to the finite size of the
target. Presumably, the soft gluons which will be exchanged
between the target and the quark-antiquark pair during 
the hadronization process  should lead to higher 
order effects in $v^2$ for the
same reasons as stated above. We will assume this is true for now, and will
keep this,  perhaps dubious assumption, in the back of our minds when we
confront the data.

\section{$\psi^\prime$ production}

The case of $\psi^\prime$ production is simplest to analyze since there
are no states below open charm threshold which contribute to its indirect
production rate. Many of the arguments discussed in this simple
case will apply for the other states as well.

The production of $\psi^\prime$ in the singlet channel begins
at $O((\frac{\alpha}{\pi})^3 v^3)$ due to charge conjugation, while the octet
channel production is $O((\frac{\alpha}{\pi})^2v^7)$. Given that numerically,
$\alpha(4m_c^2)\propto v^2$, it would seem that octet production should
be of the same order as  the singlet channel. However, the singlet cross
section vanishes at threshold where there is small $x$ enhancement
due to the gluon distribution functions and, as such, the octet actually
dominates the singlet. 

Before going on to quantitative issues however, we must address the issue of the
proper choice of the hadron mass. It is clear that the short
distance coefficients should be calculated using the quark mass, since	
binding effects are neglected by definition in this part of the
calculation. However, at face value
it seems that as far as the phase space boundary is concerned
we should be using the hadron mass instead of twice the quark mass.
This issue was vehemently debated during the workshop. As I emphasized
then, the proper choice of mass in the phase space boundary
is indeed the $2m_c$. This is the choice which is consistent with
the $v^2$ expansion, as the binding effects are always higher order in
this expansion parameter. This is best illustrated in the case of
heavy-light meson decay where the expansion parameter is $\Lambda_{QCD}/m_Q$.
In this case, we may perform an operator product expansion, with 
no question as to the proper choice of mass. At leading order in the
OPE, we find that the phase space is dictated unambiguously by the quark mass as
a consequence of unitarity. If we consider the lepton spectrum, then
there is an explicit factor of $\theta(1-2E_l/m_Q)$ in the differential rate. 
The fact that the
true phase space boundary is determined by the meson mass is seen when
one goes to higher order in the OPE where the expansion looks like
\beq
\frac{d\Gamma}{dE_l}\propto \theta(1-2E_l/m_Q)+\delta(1-2E_l/m_Q)+
\delta^\prime (1-2E_l/m_Q)+ .~.~.~.
\eeq
We see that the expansion breaks down near the partonic endpoint 
$E_l=2m_Q$ and is
signaling the need for a resummation of the non-perturbative effects.
Such a resummation leads to the construction of a structure function
which has support all the way to the hadronic endpoint.
In our case a resummation of higher order $v^2$ effects will lead to
shifting the space limits from being partonic to hadronic. Whether or not
the non-perturbative corrections are large depends upon the observable
of interest. In the example above we may safely use the partonic mass
if we are not interested in the endpoint region. With that said, let us
confront the theory with the data.

In previous analyses, performed
using the color singlet model, it was
found that the theoretical predictions needed a K factor
of 25 \cite{SCH95}. However, this discrepancy is greatly reduced once
we use twice the charm quark mass instead of the hadron mass. Indeed, given the
uncertainty in the charmed quark mass, varying $m_c$ between
$1.3$-$1.7~GeV$, changes the total cross section by a factor of 8 at 
$\sqrt{s}=30~GeV$.
This large variation is a consequence of the steep rise in the
gluon distribution at small $x$. Nonetheless, let us press on assuming
that the expansion is well behaved and consider the consequences.
In ref \cite{BRI}, it was found that, using $m_H=2m_c=3~GeV$ the color
singlet contribution fell a factor of 3 below the data. Including the 
color octet contribution leads to a fit of  the data with the choice
$\Delta_8=5.2\cdot10^{-3}~GeV$.  At large $p_T$, the authors of
\cite{CHO95} found $\langle {\cal O}^H_8 ({}^1 S_0) \rangle + 
\frac{3}{m_Q^2}\langle {\cal O}^H_8 ({}^3 P_0) \rangle=1.8\cdot 10^{-2}~GeV^3$.
If we assume $\langle {\cal O}^H_8 ({}^1 S_0) \rangle\simeq
\langle {\cal O}^H_8 ({}^3 P_0) \rangle/m_c^2$, the fixed target value
is a factor of four smaller than those found in \cite{CHO95}.
This discrepancy should not concern us, nor should we consider this
particular observable a good test of the color octet mechanism for the
reasons discussed above.

\section{$J/\psi$ production}

Using the data from proton beam fixed target experiments we may again
fit the data using the value 
\beq
\Delta_8(J/\psi)=3.0\cdot 10^{-2}.
\eeq
As in the   previous case this observable is very sensitive to the
choice of the quark mass and, as such, the fact that 
the color singlet contribution is below the data is not strong evidence
for the existence of the octet channel.
However, in this case   we may also look 
at the ratio for the direct to total cross section which is not
sensitive to the quark mass. Indeed, we find that the
pure singlet contribution gives a ratio of 0.21 whereas inclusion
of the octet with the matrix element extracted above gives\cite{BRI} 0.63
which is in much better agreement with the experimentally
extracted value of $.62\pm 0.04$ for a proton beam at $\sqrt{s}=23.7~GeV$
\cite{ANT93}. Note that this is not a trivial consequence of
fitting the color octet matrix element since the indirect contribution
is dominated by color singlet gluon fusion and the singlet matrix elements
are fixed in terms of wavefunctions \cite{EQ}.

Another very interesting observable is $\sigma_{\chi_1}/\sigma_{\chi_2}$ which
has been measured in proton as well as pion beam experiments.
The singlet cross section for $\chi_1$ production is suppressed by  a factor
of $\alpha$ relative to $\chi_2$ cross section, while the leading
octet contribution is $(O(\alpha^2v^3)$ but  is suppressed because it
is a quark initiated process. 
$\chi_2$ production on the other hand, is dominated
by color singlet gluon fusion at $O(\alpha^2v^5)$.
 The E771 experiment measured a value
\cite{hag}$\sigma_{\chi_1}/\sigma_{\chi_2}= 0.34 \pm 0.16$ for a proton 
beam at $\sqrt{s}=38.7~GeV$, and NRQCD predicts a value of 0.07\cite{BRI}\footnote{In ref. \cite{BRI} the ratio was weighted by the branching ratios to
$J/\psi$.}.
However, the relativistic corrections can be substantial\cite{srid96}
given that the leading singlet contribution scales like $\alpha^3v^3$ and
numerically $\alpha(2m_c)/\pi\propto v^3$. Indeed at $O(\alpha^2v^9)$ there
will contributions coming from intermediate $^1S_0$ and $^3P_J$ octet
as well as singlet states, as well a octet $^3D_J$ states. Though
these states are suppressed in $v^2$ given the large number of channels
which contribute, the net contribution could be substantial.
A naive use of the velocity scaling rules leads to $\simeq 0.3$\footnote{The value quoted
in the first reference in \cite{srid96} was not weighted by branching 
fractions of $\chi_J$ into 
$J/\psi$, 
contrary to what is stated in the text.}.

For pion beams
theory does not seem to do as well. 
The latest reported value for this ratio for pion beams is given
by $0.57\pm0.18$, whereas the leading NRQCD prediction is given by 
$0.07$ even including the next order corrections in $v^2$, it is clear that
the theory falls short.
Furthermore, the overall normalization for the total $J/\Psi$  and 
$\psi^\prime$ production cross sections, found using the fitted 
values of the octet matrix elements using the data from 
proton beam experiments,
 also falls short as is shown in figures 1 and 2\footnote{
data points in the plot for figure 2 in  ref. \cite{BRI} are off set
in $\sqrt{s}$ due to an error in the plotting routine.}.
\begin{figure}[t]
   \vspace{0cm}
   \epsfysize=8cm
   \epsfxsize=8cm
   \centerline{\epsffile{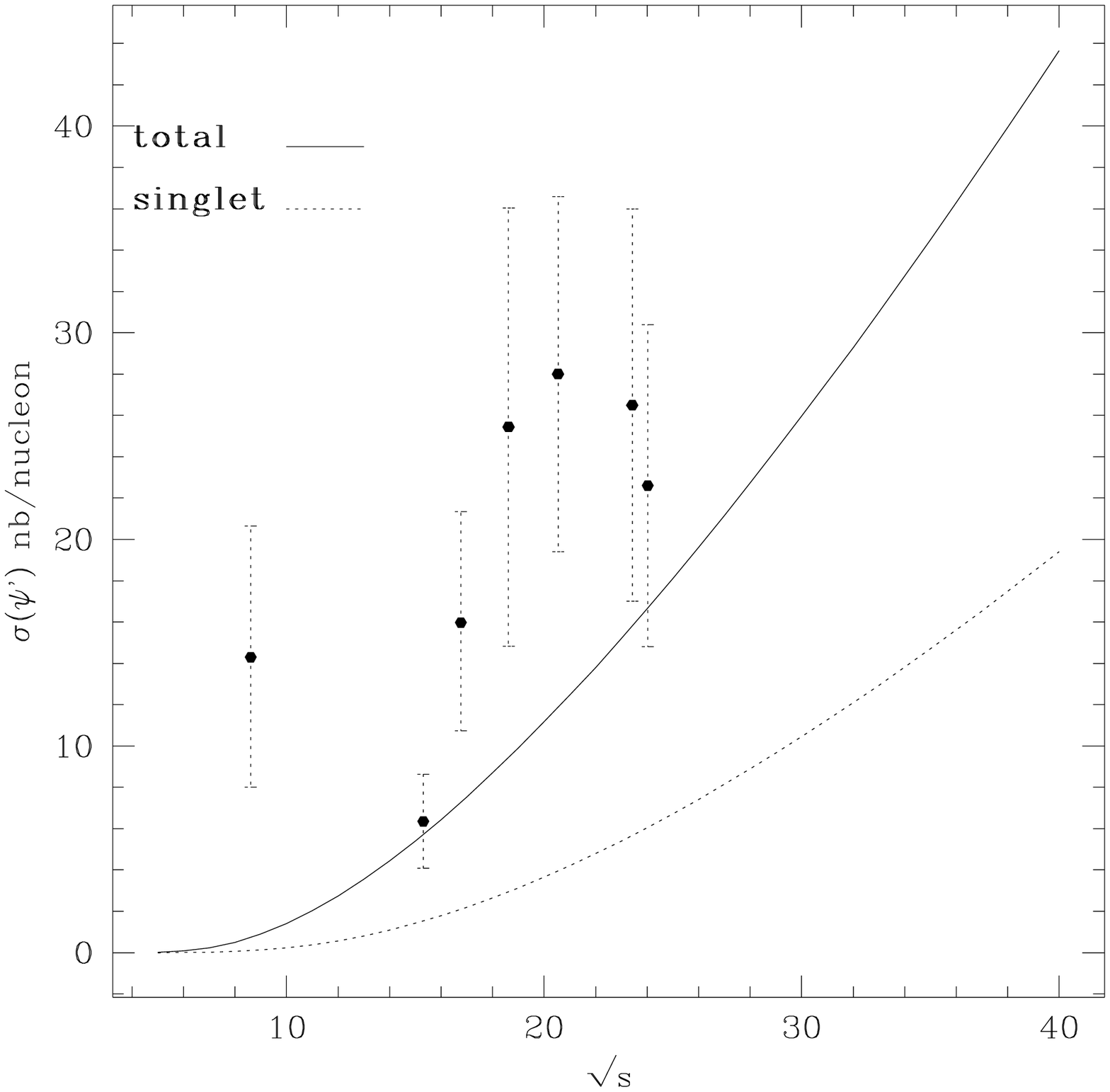}}
   \vspace*{0cm}
\fcaption{\label{pripifig} Total (solid) and singlet only (dotted) 
$\psi^\prime$ production cross section in pion-nucleon collisions 
($x_F>0$ only). The solid line is obtained 
with $\Delta_8(\psi')=5.2\cdot 10^{-3}\,$GeV${}^3$ determined using the
data from proton beam experiments.} 
\end{figure} 
\begin{figure}[t]
   \vspace{0cm}
   \epsfysize=8cm
   \epsfxsize=8cm
   \centerline{\epsffile{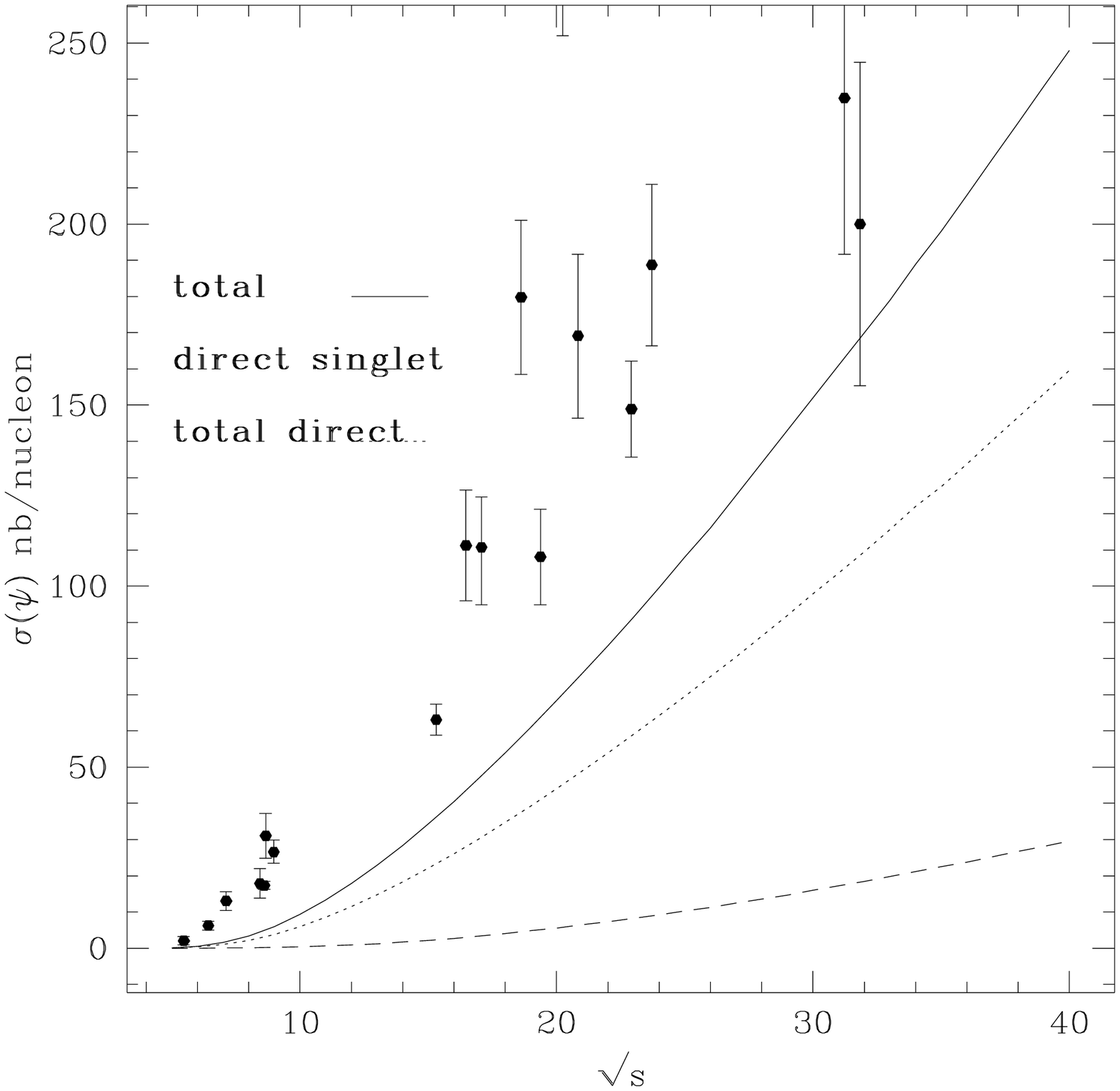}}
   \vspace*{0cm}
\fcaption{\label{psipifig} $J/\psi$ production cross sections in 
pion-nucleon collisions for $x_F>0$. Direct $J/\psi$ production 
in the CSM (dashed line) and after inclusion of color-octet processes 
(dotted line). The total cross section (solid line) includes radiative 
feed-down from the $\chi_{cJ}$ and $\psi'$ states. 
The solid line is obtained 
with $\Delta_8(J/\psi)=3.0\cdot 10^{-2}\,$GeV${}^3$.}
\end{figure}

\section{Polarization in fixed target experiments}

There are some interesting theoretical issues involving polarized
cross sections within the NRQCD formalism. However, due to space
limitations, they will not be discussed here, and I refer the
reader to refs. \cite{BRII,BRI,BRAII} for discussions. 
As we will see polarized production is a useful tool for 
investigating the octet mechanism.

Polarization measurements have been performed for both $\psi$ 
\cite{AKE93} and 
$\psi^\prime$ \cite{HEI91} production 
in pion scattering fixed target experiments. 
Both experiments observe an essentially flat angular distribution in 
the decay $\psi\to \mu^+ \mu^-$ ($\psi= J/\psi,\psi'$), 

\begin{equation}
\frac{d\sigma}{d\cos\theta }\propto 1+ \alpha \cos^2 \theta\,,
\end{equation}

\noindent where the angle $\theta$ is defined as the angle between 
the three-momentum vector of the positively charged muon and 
the beam axis in the rest frame of the quarkonium. The observed values 
for $\alpha$ are $0.02\pm 0.14$ for $\psi'$, measured at 
$\sqrt{s}=21.8\,$GeV in the region $x_F>0.25$ and 
$0.028\pm 0.004$ for $J/\psi$ measured at $\sqrt{s}=15.3\,$GeV 
in the region $x_F>0$. In the CSM, the $J/\psi$'s are predicted to be 
significantly transversely polarized \cite{VAE95}, in conflict with 
experiment.

The polarization yield of color octet processes can be calculated 
along the lines of the previous subsection. We first concentrate on 
$\psi'$ production and define $\xi$ as the fraction of longitudinally 
polarized $\psi'$. It is related to $\alpha$ by 

\begin{equation}
\alpha=\frac{1-3\xi}{1+\xi}\,.
\end{equation}

\noindent For the different intermediate quark-antiquark 
states we find the following ratios of longitudinal to transverse 
quarkonia:

\begin{equation}
\addtolength{\arraycolsep}{0.3cm}
\begin{array}{ccc}
{}^3 S_1^{(1)} & 1:3.35 & \xi=0.23\\
{}^1 S_0^{(8)} & 1:2 & \xi=1/3\\
{}^3 P_J^{(8)} & 1:6 & \xi=1/7\\
{}^3 S_1^{(8)} & 0:1 & \xi=0
\end{array}
\end{equation}

\noindent where the number for the singlet process (first line) has been 
taken from \cite{VAE95}\footnote{This number is $x_F$-dependent and we 
have approximated it 
by a constant at low $x_F$, where the bulk data is obtained from. 
The polarization fractions for the octet $2\to 2$ parton processes 
are $x_F$-independent.}. Let us add the following remarks: 

(i) The ${}^3 S_1^{(8)}$-subprocess yields pure transverse 
polarization. Its contribution to the total polarization is not 
large, because gluon-gluon fusion dominates the total rate.

(ii) For the ${}^3 P_J^{(8)}$-subprocess $J$ is not specified, 
because interference between intermediate states with different 
$J$ could occur as discussed in the previous subsection. As it 
turns out, interference does in fact not occur at leading order 
in $\alpha_s$, because the only non-vanishing short-distance 
amplitudes in the $J J_z$ basis are $00$, $22$ and 
$2(-2)$, which do not 
interfere.

(iii) The ${}^1 S_0^{(8)}$-subprocess yields unpolarized quarkonia. 
This follows from the fact that the NRQCD matrix element is 

\begin{equation}
\label{above}
\langle 0|\chi^\dagger T^A\psi\,{a_{\psi'}^{(\lambda)}}^\dagger 
a_{\psi'}^{(\lambda)}\,\psi^\dagger T^A\chi|0\rangle =\frac{1}{3}\,
\langle {\cal O}_8^{\psi'}({}^1 S_0)\rangle\,,
\end{equation}

\noindent independent of the helicity state $\lambda$. At this point, 
we differ from \cite{TAN95}, who assume that this channel results 
in pure transverse polarization, because the gluon in the 
chromomagnetic dipole transition ${}^1 S_0^{(8)}\to {}^3 S_1^{(8)}+g$ 
is assumed to be transverse. However, one should keep in mind that 
the soft gluon is off-shell and interacts with other partons with unit 
probability  prior to  hadronization. The NRQCD formalism 
applies only to inclusive quarkonium production. Eq.~(\ref{above}) 
then follows from rotational invariance.

(iv) Since the ${}^3 P_J^{(8)}$ and ${}^1 S_0^{(8)}$-subprocesses 
give different longitudinal polarization fractions, the $\psi'$ 
polarization depends on a combination of the matrix elements 
$\langle {\cal O}_8^{\psi'} ({}^1 S_0)\rangle$ and 
$\langle {\cal O}_8^{\psi'} ({}^3 P_0)\rangle$ which is different 
from $\Delta_8(\psi')$. 

To obtain the total polarization the various subprocesses have to be 
weighted by their partial cross sections. We define 

\begin{equation}
\delta_8(H)=\frac{\langle {\cal O}_8^{H} ({}^1 S_0)\rangle}
{\Delta_8(H)}
\end{equation}

\noindent and obtain 

\begin{eqnarray}
\xi &=& 0.23\,\frac{\sigma_{\psi'}({}^3 S_1^{(1)})}{\sigma_{\psi'}} + 
\left[\frac{1}{3}\delta_8(\psi')+\frac{1}{7} (1-\delta_8(\psi'))\right] 
\frac{\sigma_{\psi'}({}^1 S_0^{(8)}+{}^3 P_J^{(8)})}{\sigma_{\psi'}}
\nonumber\\
&=& 0.16+0.11\,\delta_8(\psi')\,,
\end{eqnarray}

\noindent where the last line holds at $\sqrt{s}=21.8\,$GeV (The 
energy dependence is mild and the above formula can be used with 
little error even at $\sqrt{s}=40\,$GeV). Since $0<\delta_8(H)<1$, we
have $0.16<\xi<0.27$ and therefore

\begin{equation}
0.15 < \alpha < 0.44\,.
\end{equation}

\noindent In quoting this range we do not attempt an estimate of 
$\delta_8(\psi')$. Note that taking the Tevatron and fixed target 
extractions of certain (and different) combinations of 
$\langle {\cal O}_8^{\psi'} ({}^1 S_0)\rangle$ and 
$\langle {\cal O}_8^{\psi'} ({}^3 P_0)\rangle$ seriously 
(see Sect.~5.1), a large value of $\delta_8(\psi')$ 
and therefore low $\alpha$ would be favored. Within large errors, 
such a scenario could be considered consistent with the 
measurement quoted earlier. From a theoretical point of view, however, 
the numerical violation of velocity counting rules implied by 
this scenario would be rather disturbing.

In contrast, the more accurate measurement of polarization for 
$J/\psi$ leads to a clear discrepancy with theory. In this case, we
have to incorporate the polarization inherited from decays of 
the higher charmonium states $\chi_{cJ}$ and $\psi'$. This task 
is simplified by observing that the 
contribution from $\chi_{c0}$ and $\chi_{c1}$ 
feed-down is (theoretically) small as is the octet contribution 
to the $\chi_{c2}$ production cross section. On the other hand, the 
gluon-gluon fusion process produces $\chi_{c2}$ states only in 
a helicity $\pm 2$ level, so that the $J/\psi$ in the subsequent 
radiative decay is completely transversely polarized. 
Weighting all subprocesses by their partial cross section 
and neglecting the small $\psi'$ feed-down, we arrive at 

\begin{equation}
\xi = 0.10 + 0.11\,\delta_8(J/\psi)
\end{equation}

\noindent at $\sqrt{s}=15.3\,$GeV, again with mild energy dependence.
This translates into sizeable transverse polarization

\begin{equation}
0.31 < \alpha < 0.63\,.
\end{equation}

\noindent The discrepancy with data could be ameliorated if the observed
number of $\chi_{c1}$ from 
feed-down were used instead of the theoretical value. However, 
we do not know the polarization yield of whatever mechanism is 
responsible for copious $\chi_{c1}$ production.

Thus, color octet mechanisms do not help to solve the 
polarization problem and 
one has to invoke a significant higher-twist contribution as 
discussed in \cite{VAE95}. To our knowledge, no specific mechanism 
has yet been proposed that would yield predominantly longitudinally 
polarized $\psi'$ and $J/\psi$ in the low $x_F$ region which dominates 
the total production cross section. One might speculate that both 
the low $\chi_{c1}/\chi_{c2}$ ratio and the large transverse polarization
follow from the assumption of transverse gluons in the gluon-gluon fusion 
process, as inherent to the leading-twist approximation. If gluons 
in the proton and pion have 
large intrinsic transverse momentum, as suggested by the 
$p_t$-spectrum in open charm 
production, one would be naturally led to higher-twist effects that 
obviate the helicity constraint on on-shell gluons.

\section{What Needs to Be Done}

The uncertainties in the theoretical prediction at fixed target 
energies are substantial and preclude a straightforward test 
of universality of color octet matrix elements by comparison with 
quarkonium production at large transverse momentum. Small-$x$, 
as well as kinematic effects, could bias 
the extraction of these matrix elements in different directions 
at fixed target and collider energies\cite{BRI}. The large uncertainties 
involved, especially due to the charm quark mass, could hardly 
be eliminated by 
a laborious calculation of $\alpha_s$-corrections to the 
production processes considered here. 
To more firmly establish existence of 
the octet mechanism there are several experimental
measurements which need to be performed. Data on polarization is presently
only available for charmonium production in pion-induced collisions. A 
polarization measurement for a proton beam would be very interesting
given that we seem to have a better handle on the theory in this case,
as is demonstrated by the observed value of the $\chi_1/\chi_2$ ratio.
A measurement of polarization at large transverse momentum or 
for bottomonium is of crucial importance, 
because higher twist effects should be suppressed. 
Furthermore, a measurement of direct
and indirect production fractions in the bottom system would provide further 
confirmation of the color octet picture and constrain the color octet 
matrix elements for bottomonium. 

From a theoretical standpoint there are still several issues that
warrant further investigation. To begin with, factorization 
in hadro-production of quarkonia is presently just a working hypothesis. This
is true at large $p_T$ as well as at small $p_T$.
Of course while formally this puts both calculation on the
same theoretical footing; practically, there is still an important
difference between the two cases. Namely, the size of the
higher twist effects will be suppressed by $1/p_T^2$ at large
$p_T$, as opposed to $1/(4m_c^2)$, for our fixed target cross section.
It is furthermore possible, and this again applies to the case
of large $p_T$ as well, that the higher twist effects could be
enhanced by powers of $1/v$.
Indeed, understanding higher twist effects in quarkonium production
is complicated by the presence of the 
scales $mv$ and $mv^2$.  Needless to say, there is still much work to
be done on this subject before we can get a good handle on the
errors due to higher twist effects.

\nonumsection{Acknowledgments}

I wish to thank Martin Beneke for his
collaboration on this subject, as well as Rick Jesik Victor
Koreshev and K. Hagan for useful discussions. 
 Finally, I wish to thank the organizers
of the workshop for their hospitality.

\nonumsection{References}

\end{document}